\documentclass[modern]{aastex62}

\usepackage{acro}
\usepackage{amsmath}
\usepackage{color}

\begin{document}

\title{Evidence for an ancient sea level on Mars}

\author[0000-0001-8864-4748]{Abbas Ali Saberi}
\affiliation{Department of Physics, University of Tehran, P. O. Box 14395-547, Tehran, Iran}
\affiliation{Institut f\"ur Theoretische
	Physik, Universit\"at zu K\"oln, Z\"ulpicher Str. 77, 50937 K\"oln,
	Germany}
\email{ab.saberi@ut.ac.ir \& ab.saberi@gmail.com}

\begin{abstract}
  Mars shares many similarities and characteristics to Earth including various geological features and planetary structure. The remarkable bimodal distribution of elevations in both planets is one of the most striking global features suggesting similar geodynamic processes of crustal differentiation on Earth and Mars. There also exist several evidences, based on geographic features resembling ancient shorelines, for existence of an ancient martian ocean in the northern hemisphere which covers nearly one third of the planet's surface. However, the interpretation of some features as ancient shorelines has been thoroughly challenged that left the existence of a primordial martian ocean controversial. Moreover, if oceans were formerly present on Mars, there is still a big ambiguity about the volume of water with the estimations ranging over $4$ orders of magnitude. Here we map the martian sea level problem onto a percolation model that provides strong evidence that the longest iso-height line on Mars that separates the northern and southern hemispheres, acts as a critical level height with divergent correlation length and plays the same role as the present mean sea level does on Earth. Our results unravel remarkable similarities between Mars and Earth, posing a testable hypothesis about the level of the ancient ocean on Mars that can be answered experimentally by the future investigations and spacecraft exploration.
\end{abstract}

\section*{ }

\begin{figure}
\centerline{\includegraphics[width=0.8\columnwidth,clip=true]{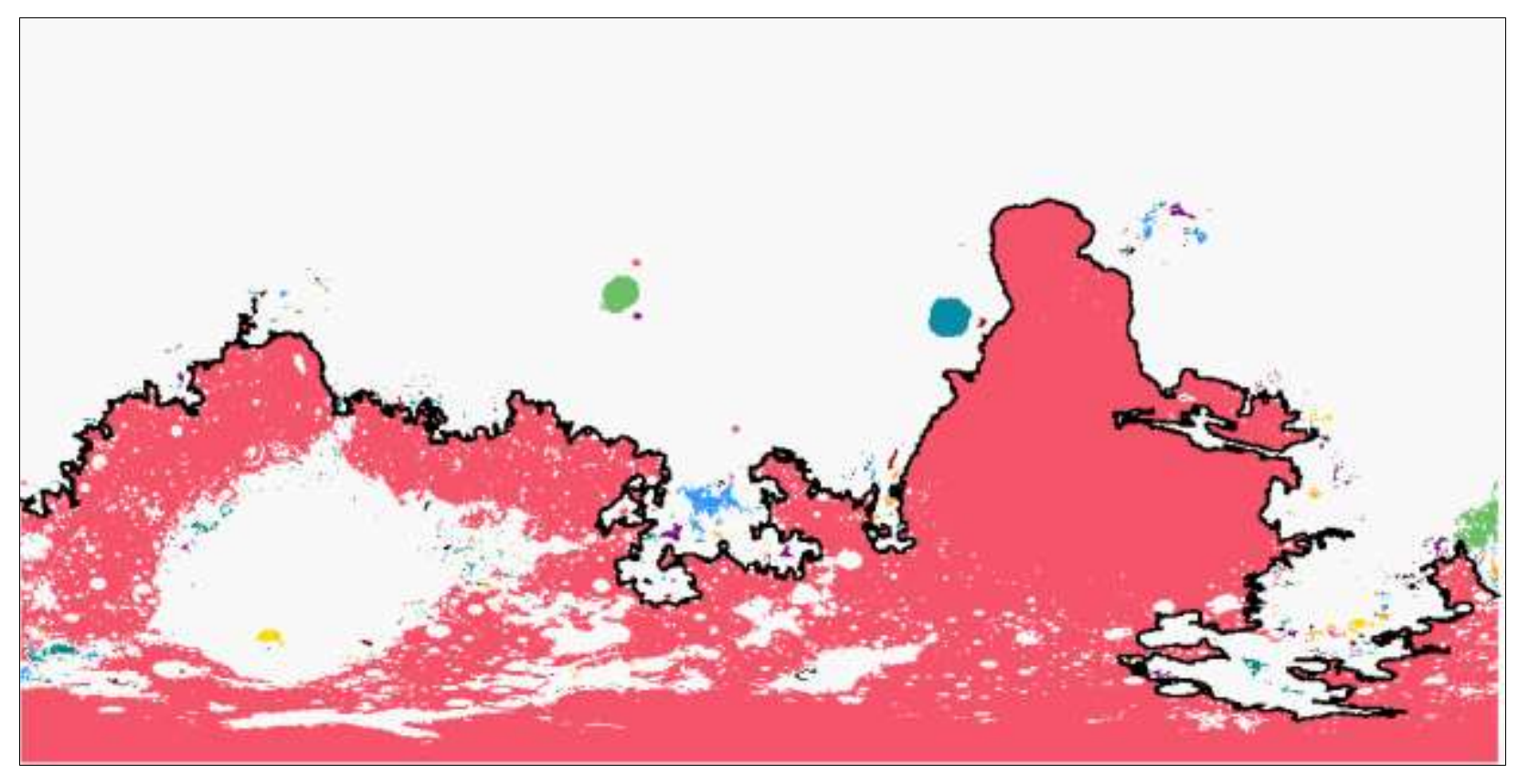}}
\caption{\label{Fig1}Two distinct southern and northern hemispheres on Mars separated by the proposed sea level (the solid line) at height $h=h_c$. Different colors show isolated islands, and regions under water are shown in white.}
\end{figure}

Global topography reflects complex history of various dynamic processes that have shaped Mars on a broad range of spatial scales and contains important encoded information of the
planet's evolution \citep{Smith1999}.
The major feature of the martian crust is that it exhibits a sharp contrast between the northern and southern hemispheres, known as the martian dichotomy~\citep{Carr1981, Mutch1976} (Fig. \ref{Fig1}).
This hemispheric dichotomy is manifest topographically as the presence of a heavily cratered and elevated highlands in the southern hemisphere distinguished from the lowland plains in the northern hemisphere by a reduction in crustal thickness of $\sim $30 km \citep{Smith1996, Zuber2001}. The origin of the hemispheric dichotomy is generally unknown, but some hypotheses for its possible explanation include a single mega-impact or impacts into the northern hemisphere \citep{Wilhelms1984}, 
thinning of the northern-hemisphere crust by regionally
vigorous mantle convection \citep{Lingenfelter1973}, a differential-rotation mechanism excited by a one-plume mantle convection \citep{zhong2009migration}, and an early period of tectonic-plate recycling \citep{Sleep1994}. It has also been suggested that the northern hemisphere was
once covered by a vast ocean \citep{Baker1991, Perron2007}. The best evidence for existence of an ancient ocean on Mars mainly relies on the observation of several possible palaeoshorelines
that distribute predominantly near the margins of the northern plains of Mars \citep{Baker1991, Parker1989, Parker1993, Clifford2001}. Two of these river-valley-like patterns can even be traced without major interruptions for thousands of kilometers. However, these notions have been thoroughly challenged by \cite{Malin1999} based on the observation of elevation variations of significant long-wavelength trends along the identified shorelines of up to a couple
of kilometers which indeed do not follow surfaces of equal gravitational potential that should hold for a sea level \citep{Malin1999, Head1998, Head1999, Carr2003} as it does on Earth. However, further studies have continued the debate on how the palaeoshorelines could have been deformed to display significant elevation variations today \citep{di2010ancient, chan2018new, citron2018timing}.\bigskip

Despite their essential differences, Mars and Earth share striking similarities in their global topographic features with a similar bimodal nature that can be indicative of an analogous geodynamic activity of the two planets in the distant past \citep{Perotti2011}. In this context, the martian dichotomy may suggest an early oceanic and continental crustal differentiation as a consequence of plate tectonic processes, with evident differences explained by a premature end to geodynamic processes and consequent isostatic readjustment of the Mars's surface. Here we provide strong evidence in support of this hypothesis, and introduce a level height as a strong candidate for an ancient sea level on Mars which resolves the above mentioned controversies and also consistent with our observations on Earth.

\begin{figure}
\centerline{\includegraphics[width=0.8\columnwidth,clip=true]{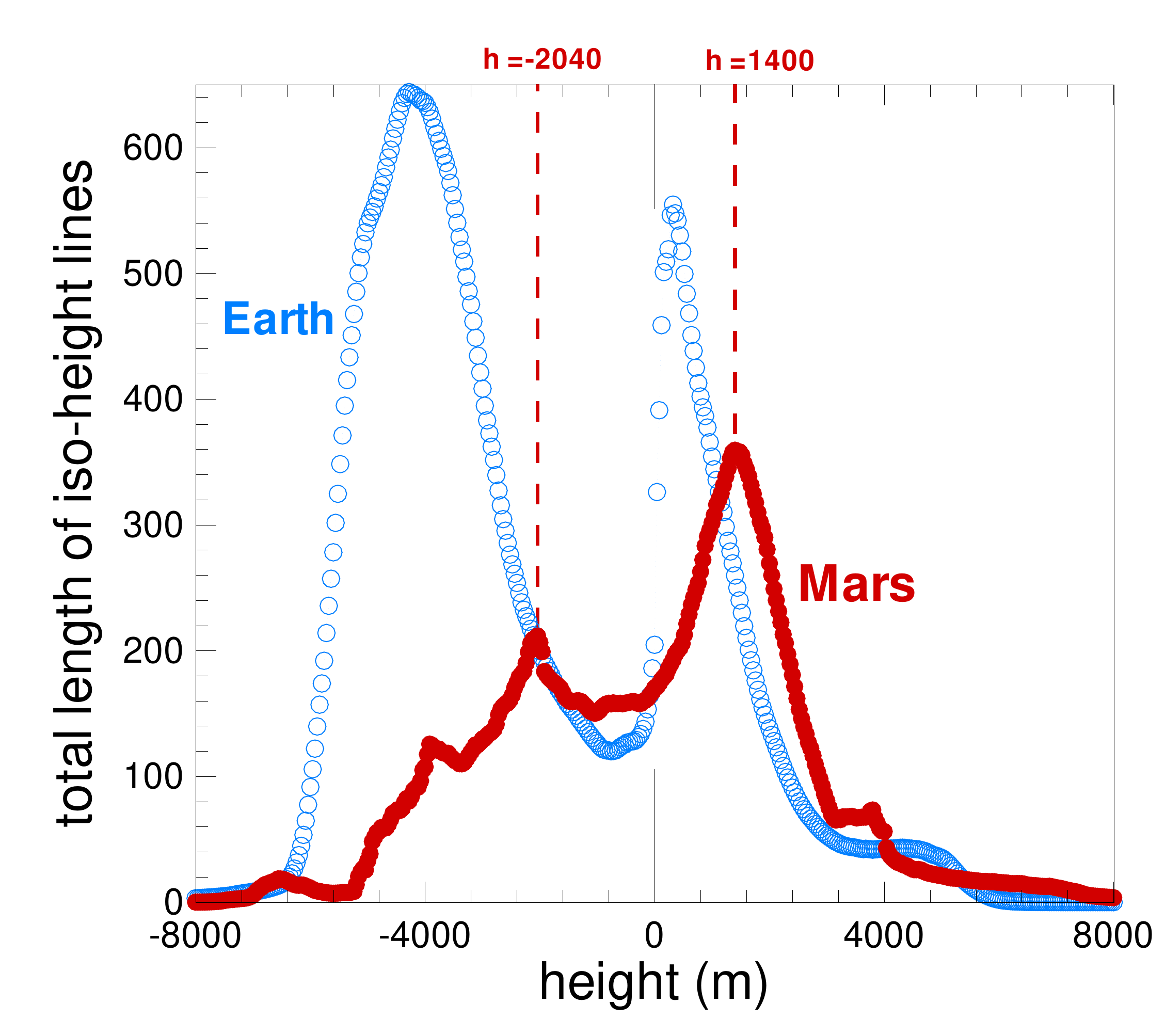}}
\caption{\label{Fig2}Total length of the iso-height lines on Mars (solid circles) and Earth (open circles) as a function of altitude. The zero level height corresponds to the present mean sea level on Earth, but has a different meaning on Mars. The two vertical dashed-lines specify two distinct level heights $h=-2040$ m and $h=1400$ m with maximum length of the iso-height lines in the northern lowlands and southern highlands of Mars, respectively. The same plot for Earth containing two oceanic and continental peaks, unravels further common similarities between these two planets. Note that the present mean sea level on Earth, marked by the vertical solid lines, is located just below the continental maximum. All length scales are expressed in units of the average radius of the corresponding planet. The resolution of the analysed data sets for Earth and Mars are 10800$\times$21600 and 4000$\times$8000 grid points, respectively. Spherical shape of the planets was taken into account when calculating the length of the iso-height lines. Details of calculations and finite-size scaling analysis at different resolutions are given in the Appendix.}
\end{figure}

A giant ancient ocean covering the entire northern lowlands on Mars, with any estimated amount of water \citep{Cart1996, Luo2017}, should have been surrounded by a relatively long gravitational equipotential contour line and should have left manifest effects on the present topography of Mars. In order to search for such effects, we use the high-resolution and globally distributed topographic data of Mars provided by the Mars Orbiter Laser Altimeter (MOLA) \citep{Zuber1992}, an instrument on the Mars Global Surveyor (MGS) spacecraft \citep{Albee1998}. 
Figure \ref{Fig2} gives an impression of how the total length of the iso-height (or contour) lines on Mars varies by altitude. Remarkably, this curve resembles the bimodal histogram of the martian topography indicating two evident peaks at level heights $h=-2040$ m and $h=1400$ m with a local minimum in between~\footnote{Zero level height on Mars is defined as the equipotential surface whose average value at the equator is equal to $3396.2$ km.}. Interestingly, a similar pattern is observed \citep{Saberi2013} on Earth with two distinct maxima in the continental platforms and oceanic floors which displays a major difference that the peak corresponding to the oceanic floor on Earth is relatively much more prominent than its twin on Mars (Fig. \ref{Fig2}). This difference can be due to the long period of geodynamic and morphological inactivity of Mars.
The striking observation is that the maximum total length of the iso-height lines on Earth is located at $\sim320$ m, just above the present mean sea level on Earth with a very sharp increase \citep{Saberi2013} (Fig. \ref{Fig2}). The sharp steepening of the curve for $0\lesssim h\lesssim 320$ m, may reflect the historical trace of past sea-level fluctuations coupled to its corresponding paleotopography over the geological timescales on Earth which is estimated by \cite{hallam1992phanerozoic} to have been in the same range of height. However, such steepening effect is absent in the same curve for the Mars (Fig. \ref{Fig2}) which may be explained by the absence of water and the geodynamic processes over a long period of time after their early existence on Mars.

\begin{figure}
	\centerline{\includegraphics[width=0.6\columnwidth,clip=true]{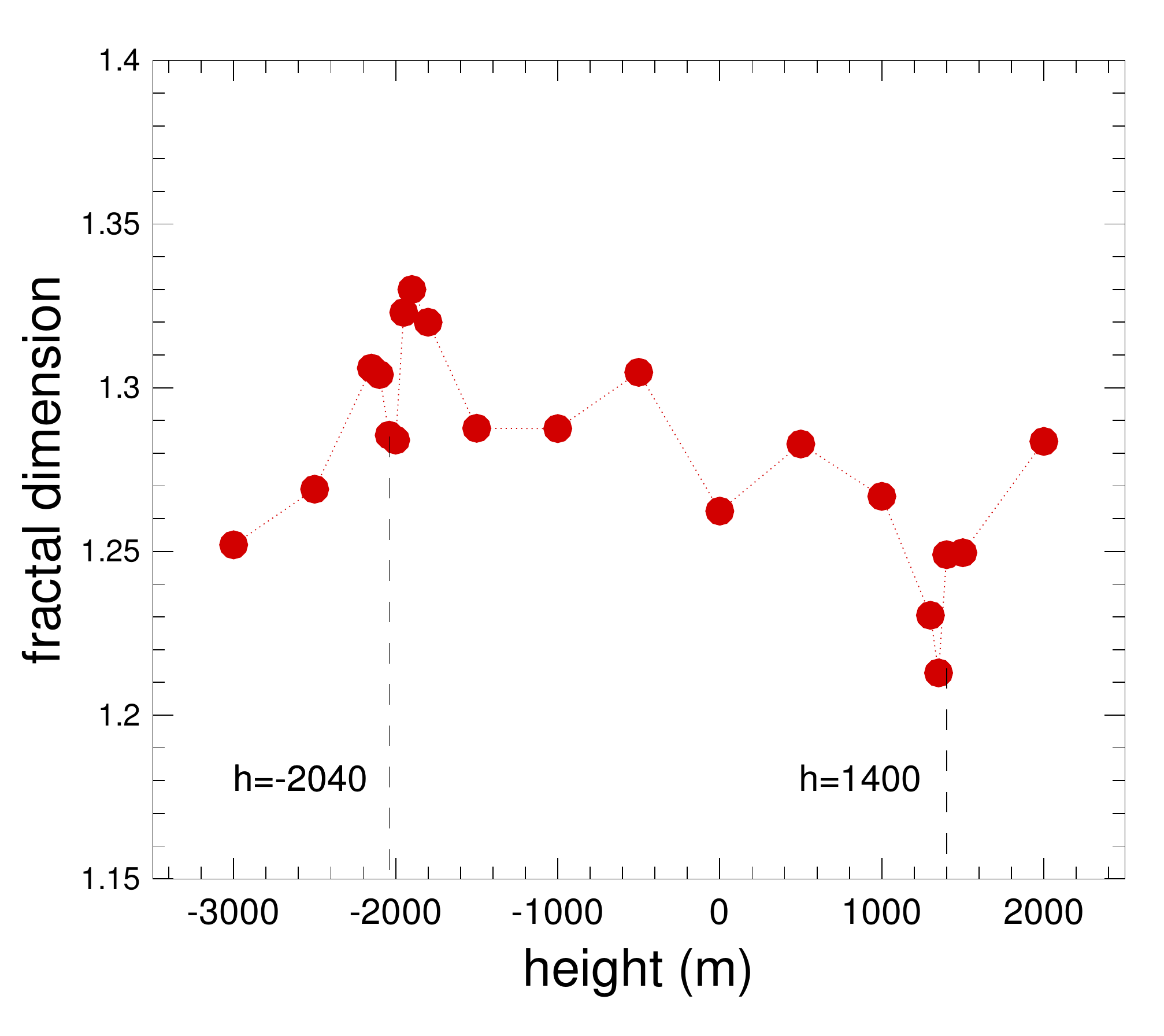} 
	} 
	\caption{\label{fracDim} Fractal dimension of the longest isoheight line on Mars as shown in Fig. \ref{Fig1} at different altitude. This is comparable with the overall most probable fractal dimension $D_f\simeq 1.2$ found for the shorelines on Earth \citep{Boffetta2008, baldassarri2008fractal}.
}\end{figure}

 Measurement of the fractal dimension of the longest isoheight line on Mars (the solid line in Fig. \ref{Fig1}) at different altitude reflects the geometrical irregularity of the martian topography with a significant response around the level height $h=1400$ m (Fig. \ref{fracDim}), which is very close to the overall most probable value $D_f\simeq 1.2$ found for the shorelines on Earth \citep{Boffetta2008, baldassarri2008fractal}. A rather similar jump in the fractal dimension $D_f$ has also been observed on Earth  around the present mean sea level at $h=0$, while there is no notion of irregularity on the Moon \citep{baldassarri2008fractal}.

In the following, we apply a quantitative framework based on percolation theory \citep{Sahimi1994, Stauffer1994, Isichenko1992, Saberi2015} to uncover further signatures in the statistical properties of the iso-height lines on Mars unraveling further pattern similarities with Earth. In this framework, a correspondence between the statistical topography of Mars and the standard percolation problem---the simplest statistical model with nontrivial critical behavior---is established in which the random contour lines are associated with percolation cluster boundaries. We assume the martian topography on a sphere of unit radius, so all length scales here are expressed in units of the Mars's average radius. One can imagine flooding this global landscape homogeneously so that the global sea level all over the planet would coincide with a spherical surface assigned by the corresponding level height depending on the amount of water (starting from the highest level to the lowest one). All parts above the water level are then colored differently as disjoint islands (or percolation clusters), and the rest is left white (Fig. \ref{Fig1}). The level height is the control parameter in this percolation picture in which the islands (clusters) can gradually grow and join each other and, if the model is critical, at some critical level $h=h_c$ there will appear a super-continent (percolating cluster) accompanied by a super-ocean containing an area of $\mathcal{O}(A)$ with $A=4\pi$ being the total area of the sphere of unit radius. We will refer to this type of flooding (top-to-bottom) as 'continental clustering'. We will also use an alternative approach starting from the lowest level of water to the highest level in which all parts below the water level are colored differently as disjoint lakes, and the rest is left white, that will be referred as 'oceanic clustering'.

\begin{figure}
	\centerline{\includegraphics[width=0.8\columnwidth,clip=true]{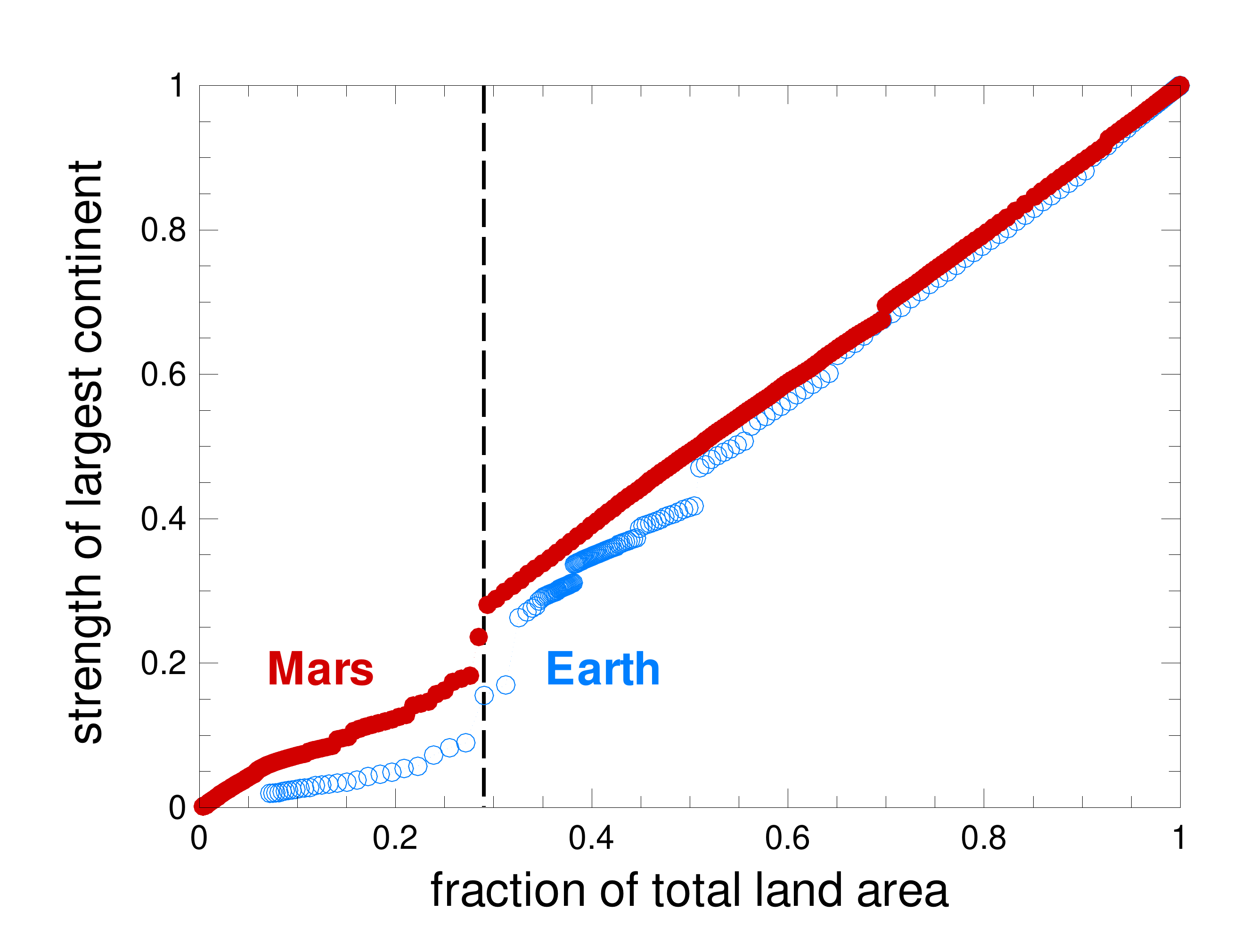}
	}
	\caption{\label{Fig3}Relative surface area of the largest island to the total area ($4\pi$) of Mars (solid circles) and Earth (open circles), as a function of the total fraction of the landmass area. The vertical dashed-line marks the percentage ($29 \%$) of the present landmass area on Earth. This also coincides with the level height $h=1400$ m on Mars including the maximum length of the iso-height lines (Fig. \ref{Fig2}).  
}\end{figure}

Figure \ref{Fig3} presents the strength of the largest continent on Mars (solid circles) as a function of the total fraction of the landmass area. We define the strength of the largest continent as the ratio of the area of the largest island to $4\pi$.  Interestingly, this quantity exhibits a major jump at a certain fraction marked by the vertical dashed-line in Fig. \ref{Fig3}, indicative of a percolation phase transition in the topography of Mars. This point is characterized by the largest gap in the strength of the largest continent, however, there are also some micro jumps due to the microtransitions discussed by~\cite{Habib2013}, \cite{Schrenk2012}, \cite{Riordan2012}, \cite{Nagler2012}, and \cite{Nagler2014}. A remarkable observation is obtained when we compare our finding for Mars with the similar quantity on Earth (shown by the open circles in Fig. \ref{Fig3}): In both planets the major geometrical phase transition occurs in the same fraction $\sim 29 \%$ of the total landmass area, which is also consistent with the present mean sea level on Earth that makes up $\sim 71\%$ of it's surface. In what follows, we present some arguments that indicate this observation can not be accidental.  

\begin{figure}
	\centerline{\includegraphics[width=0.6\columnwidth,clip=true]{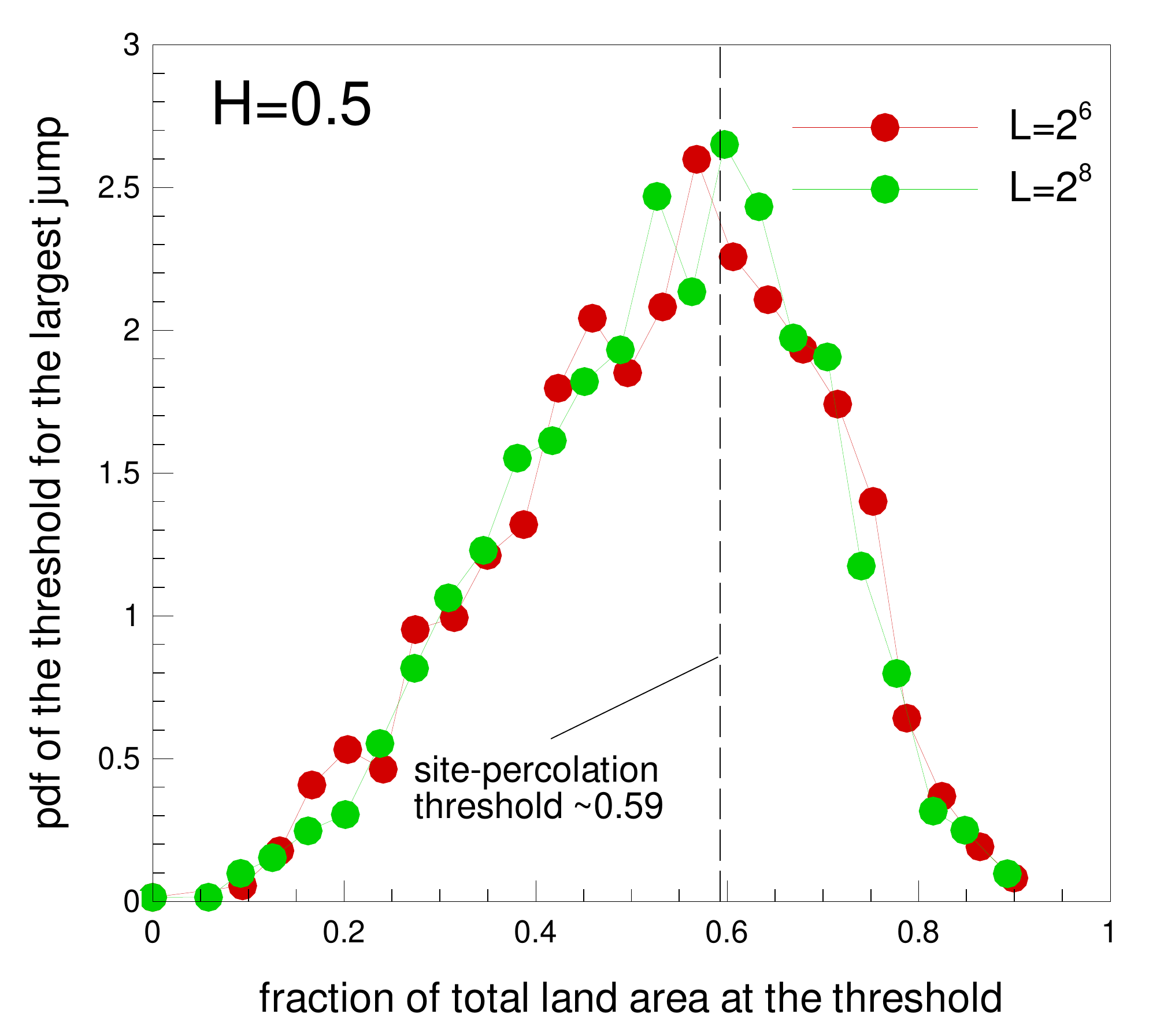}
	}
	\caption{\label{FBM-pc} The probability distribution function (pdf) of the fraction of the total land area at which the largest jump in the order parameter occurs on 2D fBm surfaces with $H=0.5$ and two different sizes $L=2^6 $ and $2^8$. The vertical dashed line shows the critical threshold for the 2D standard site-percolation problem on a square lattice.  
}\end{figure}

Let us consider the fractional Brownian motion (fBm) model of topography with a given Hurst exponent $H$, first introduced by \cite{Mandelbrot1975}. Recently, it has been shown that \citep{Landais2019} both Mars and Earth, as two samples of long-range correlated topographies, share the same Hurst exponent $H\sim 0.5$ at length scales larger than $\sim 10$ km. This allows us to generate various copies of such planet-like topographies to provide statistical comparison with our empirical observations for real topographies. To this aim, we have generated $10^6$ fBm surfaces with $H= 0.5$, and applied the percolation analysis to each height sample. We find that the strength of the largest continent shows similar behavior as shown in Fig. \ref{Fig3} with a number of jumps, with a remarkable difference that in the fBm surfaces the position of the largest jump is randomly distributed, in accord with the non-self-averaging property of these reliefs \citep{Du1996} (four sample examples are shown in Figure \ref{Fig11} of the Appendix). Figure \ref{FBM-pc} presents the probability distribution function (pdf) of the fraction of the total land area at which the largest jump occurs on fBm surfaces with $H=0.5$ on square lattice of two different sizes $L=2^6 $ and $2^8$. Remarkably, we find that the most probable fraction for the occurrence of the largest jump in the order parameter is at $\sim 0.59$, i.e., at the critical threshold of a 2D standard site-percolation on a square lattice. This suggests that (i) the largest jumps observed on Mars and Earth are tightly related to percolation-type transitions, (ii) there is a less chance to observe the largest jump at a fraction $\sim 0.29$, and (iii) it is statistically  unlikely that the two planets share the same critical fraction, which indicates that the observed coincidence of the threshold level $\sim 0.29$ for Mars and Earth can not be accidental. For the water-less Moon, we find that the critical fraction is located at $\sim 0.5$, reflecting its nearly random topographic structure---see Section B in the Appendix. The critical fraction $\sim 0.29$ on Mars corresponds to the level height $h_c\sim 1400$ m with the maximum length of the iso-height lines in Fig. \ref{Fig2}.

\begin{figure}
	\centerline{\includegraphics[width=0.8\columnwidth,clip=true]{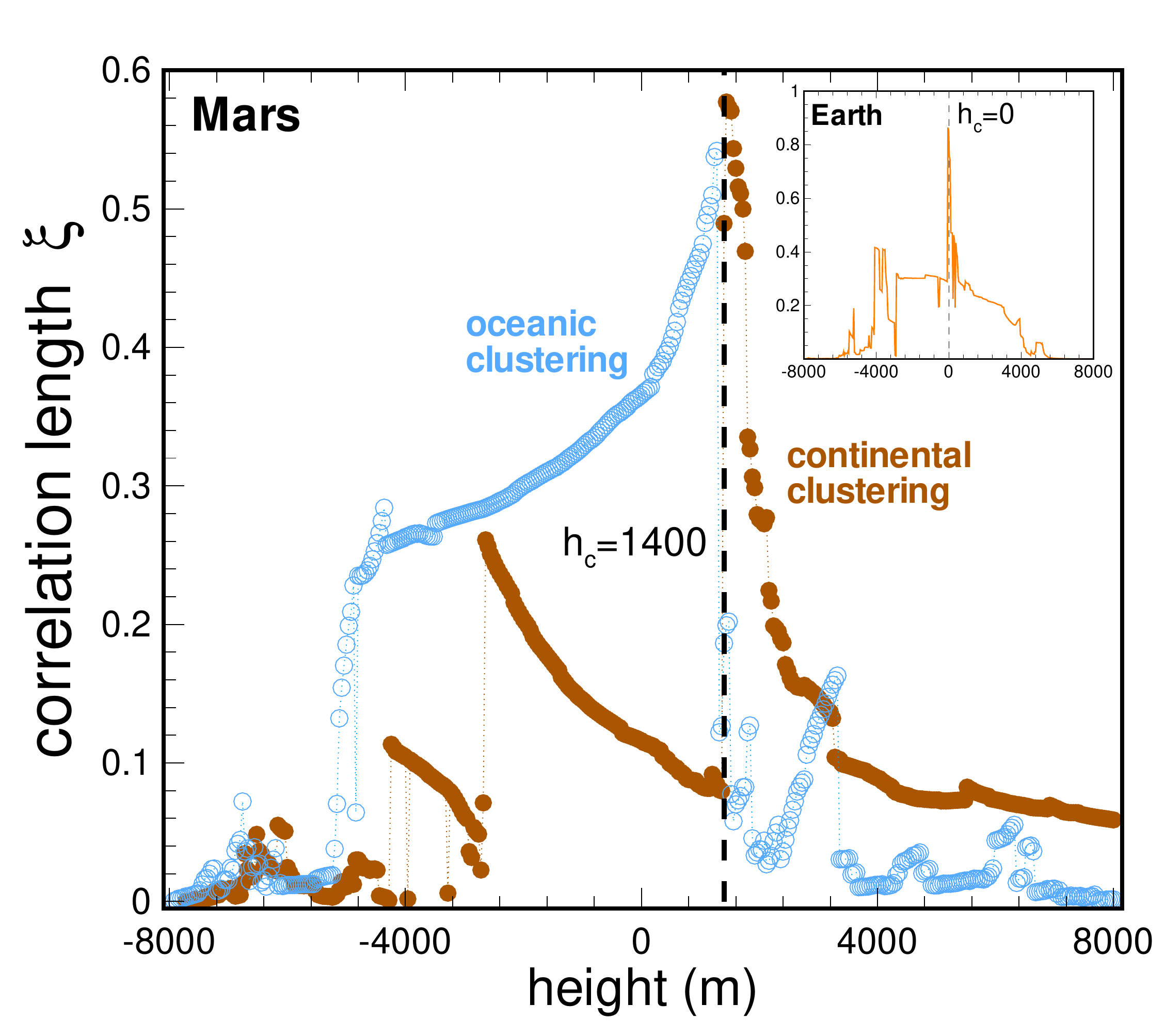}
	}
	\caption{\label{Fig4}Main: Correlation length of islands (solid circles) and oceans (open circles) obtained from the continental and oceanic clustering approaches, respectively, as a function of altitude on Mars. Both correlation lengths give rise to a remarkable divergence at the same critical level $h_c=1400$ m. Inset: Correlation length of islands on Earth diverges at the present mean sea level $h_c=0$.}\end{figure}

The other percolation observable that can help recognizing further pattern similarities between Mars and Earth is the correlation length $\xi$.
The correlation length in our percolation framework at a given height level $h$, is defined \citep{Stauffer1994} as the average distance of sites belonging to the same island, \begin{equation}\xi^2=\frac{2\sum'_s R_s^2 s^2 n_s(h)}{\sum'_s s^2 n_s(h)},\end{equation} where $n_s(h)$ denotes the average number of islands of surface area $s$ at level $h$, $R_s$ is the radius of gyration of a given island of surface area $s$, and the prime on the sums indicates the exclusion of the largest island in each measurement. We find that the correlation length for both continental and oceanic clustering diverge at the same level height $h_c=1400$ m (Fig. \ref{Fig4}). This observation becomes more striking when compared with the computed correlation length on Earth. As shown in the Inset of Fig. \ref{Fig4}, the correlation length diverges exactly at the present mean sea level on Earth which can be meaningful for Mars as well. Putting all these evidences together, we come to the main proposal of the present study that Mars' northern hemisphere was covered by an ocean early in the planet's geologic history with the water level up to the height $\sim 1400$ m.

\begin{figure}
	\centerline{\includegraphics[width=0.9\columnwidth,clip=true]{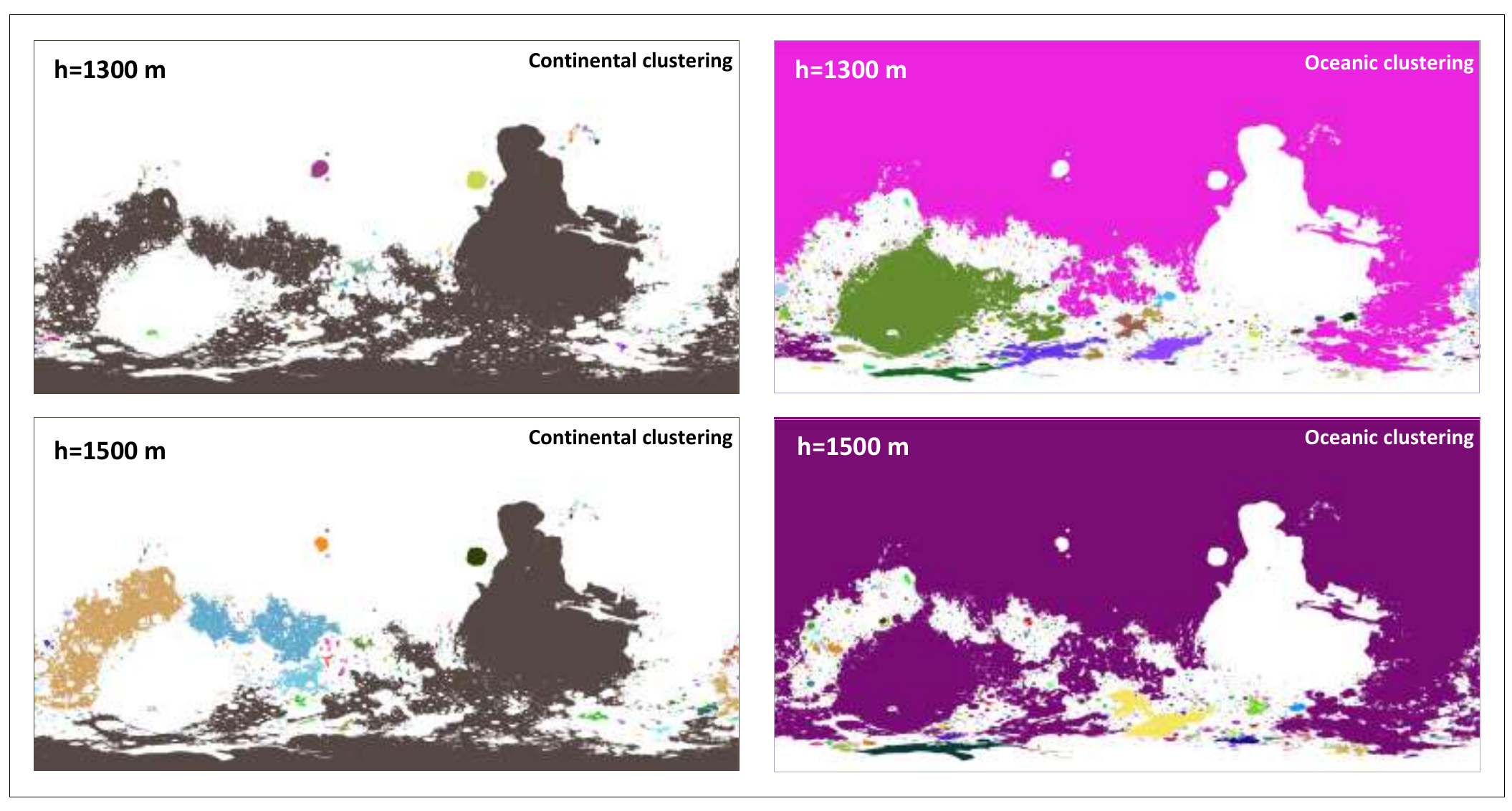}
	}
	\caption{\label{Fig5}
		Schematic illustration of the continental/oceanic aggregation (left/right column) just below (top row) and above (bottom row) the critical height level at $h_c=1400$ m. A different color is assigned to each isolated continent/ocean. This shows a remarkable percolation transition at $h_c=1400$ m around which the major parts of the landmass join together.
}\end{figure}

A close look at the continental and oceanic clusters around the critical level shows that there are three or four major islands that dominate the critical behavior (Fig. \ref{Fig5}). The first major continent includes some parts of the Terra Cimmeria and Terra Sirenum landmasses connected to the Tharsis Montes, Olympus Mons and Alba Mons (left column in Fig. \ref{Fig5}). The second and third major islands are located around the Hellas Planitia including Terra Sabaea and Hesperia Planum, respectively. Hellas Planitia is also a key region in the oceanic clustering (right column in Fig. \ref{Fig5}). Our analysis shows that it is disconnected from the global northern ocean just below the critical level and gets connected just above $h_c$.

The continental and oceanic maps (see Fig. \ref{Fig5}) may also reveal more hidden structural information on Mars when they are compared with those observed on Earth (see Fig. \ref{Fig12} of the Appendix). Specifically, when the map of disjoint islands at $h=100$ m above the present mean sea level on Earth is compared with the Earth's map of tectonic plates, one interestingly observes that every disjoint landmass is approximately surrounded by a major plate boundary (Fig. 15 in Ref. \citep{Saberi2015}). In case of the same scenario on Mars, we now have the first impression about how could the map of early plate tectonics be on Mars (Fig. \ref{Fig5}).

Our results establish an unexpected connection between the global topographic features of Mars and its ancient geological evolution and structure. On the basis of percolation theory and assuming an early history comparable to that of the Earth, we are able to address some fundamental questions about Mars, especially on the inventory and role of water in the geophysical and  geological evolution of Mars. Our analysis gives rise to a very distinct critical level height that is in all aspects comparable to the present mean sea level on Earth. This result implies the existence of long-range correlations between the iso-height sites on Mars around the observed critical level which is indicative of the presence of an ancient super ocean with the specific sea level at $\sim 1400$ m. We hope that progress in direct experimental evidence of our observations will be attained by the future investigations and spacecraft exploration that includes low-altitude platforms and in situ observations around the suggested critical level height on Mars.

\textit{Acknowledgment.} I would like to thank H. Dashti for his kind helps on programmings. Supports from the Alexander von Humboldt Foundation and the research council of the University of Tehran are acknowledged.

\bibliography{refs}

\appendix

\section{Resolution effects on total length of the iso-height lines on Mars}

\begin{figure}
	\includegraphics[width=0.8\columnwidth,clip=true]{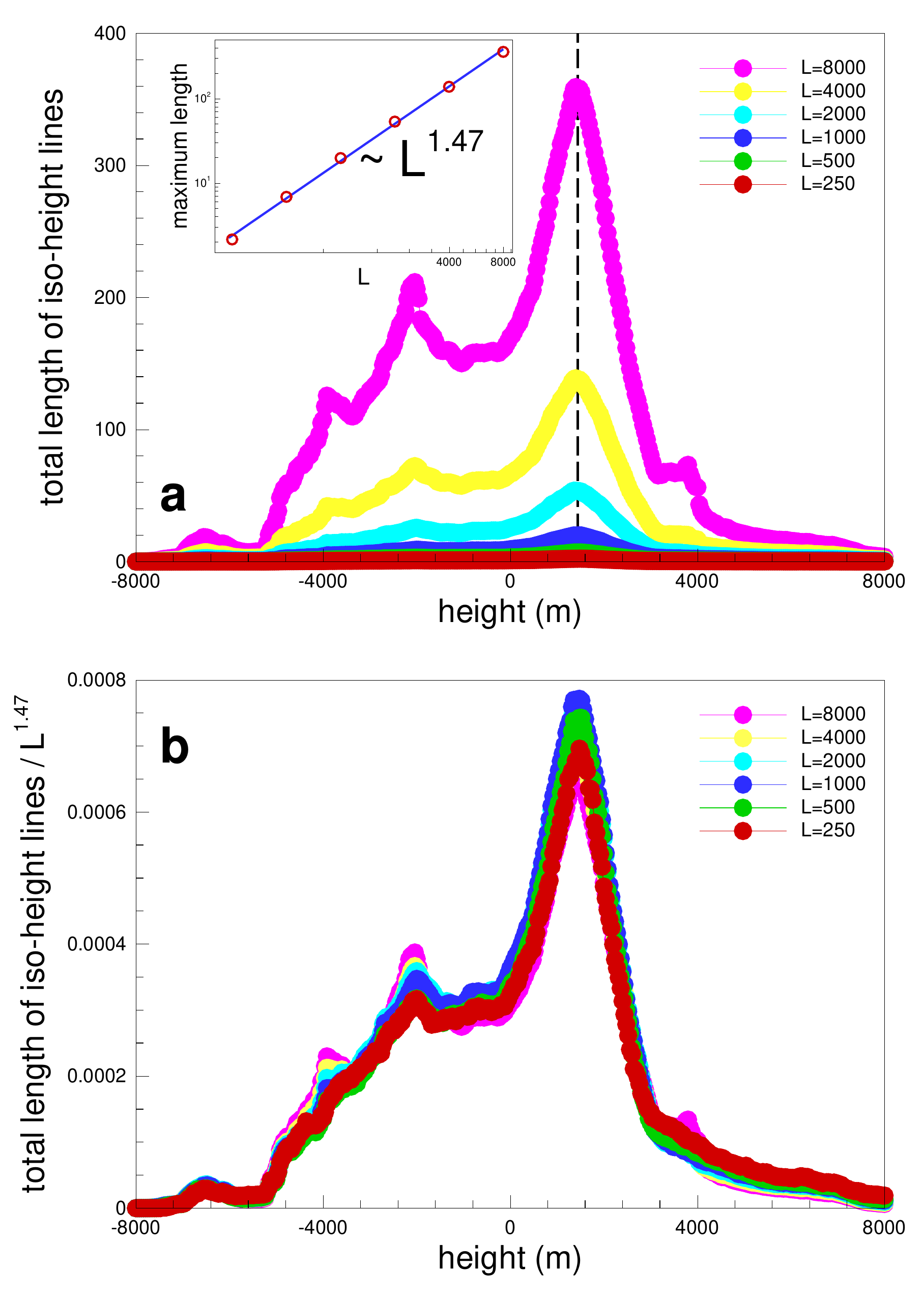}
	\caption{\label{Fig8}
		(a) Main: Total length of the iso-height lines on Mars for different resolutions. Inset: Scaling behavior of the maximum lengths at different resolutions shown by the dashed line in the main panel versus the number of grid points $L$ along the latitude. (b) Data collapse of different resolutions for the rescaled lengths of fractal dimension $D_f=1.47$.   
}\end{figure}

In this section, we study the finite size effects by altering the resolution of the grid points in martian topography. Starting from the original data set on a sphere of unit radius with 4000$\times$8000 grid points of total number $(L_0/2)\times L_0$ (with lattice spacing which is latitude-dependent in the spherical coordinates), we furnish the lattice with 2000$\times$4000 blocks each of which consists of four grid points. We then assign the average of the four heights to the center of each block which results in a  coarse-grained topography on a sphere of unit radius with total number of $(L_0/4)\times (L_0/2)$ grid points. By repeating this procedure, we can produce copies of martian topography in different resolutions $L=L_0, L_0/2, L_0/4,\cdot\cdot\cdot$. For each resolution $L$ we measure the total length of the iso-height lines which are believed \citep{Mandelbrot1982, Kondev2000} to be fractal sets of dimension $D_f=2-H$ with Hurst exponent $H$, for the level set of a
random self-affine surface. The results are shown in Figure \ref{Fig8}.

As shown in the Inset of Figure \ref{Fig8}(a), the total length $l$ of the iso-height lines exhibits a scaling relation with the number of grid points $L$ along the latitude, i.e., $l\sim L^{1.47}$. When we use the scaling relation $D_f=2-H$ for self-affine surfaces with the exponent $D_f=1.47$ to estimate the Hurst exponent for martian topography, it gives $H=0.53$ which is in perfect agreement with its recent estimation by \cite{Landais2019}. One can use the fractal dimension $D_f$ of the level sets to collapse all data for different resolutions onto a single universal curve as shown in Figure \ref{Fig8}(b).

\begin{figure}
	\includegraphics[width=0.8\columnwidth,clip=true]{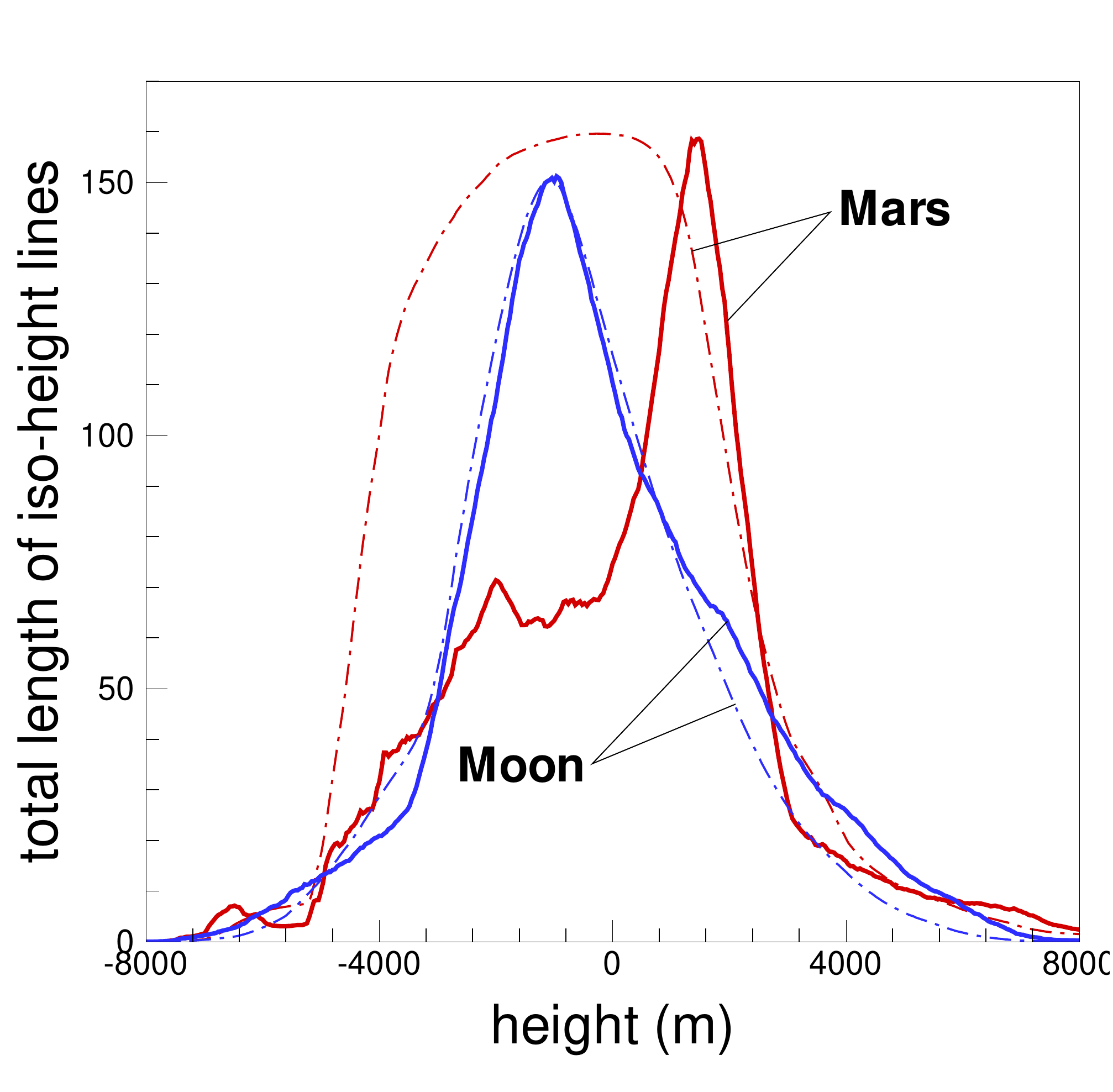}
	\caption{\label{Fig9} Total length of the iso-height lines on Mars (on a coarse grained sample of size $2000\times 4000$) and the Moon as a function of altitude computed from their original topographies (solid lines) and their randomized samples (dotted-dashed lines). Each result for the randomized samples are scaled with a constant for clarity and ease of comparison.
}\end{figure}

\begin{figure}
	\includegraphics[width=0.8\columnwidth,clip=true]{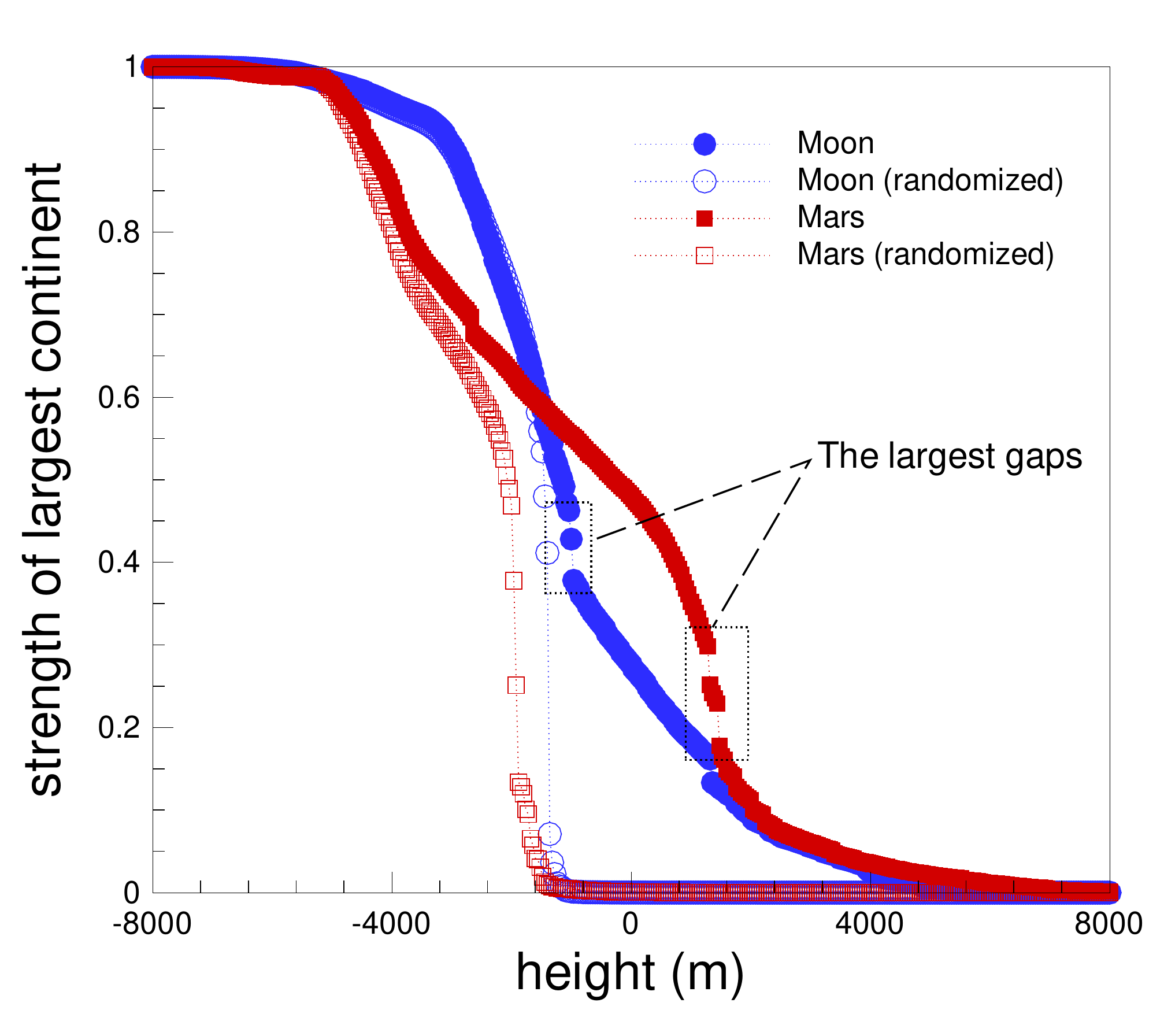}\caption{\label{Fig10}Strength of the largest continent on Mars and the Moon as a function of altitude computed from their original topographies (solid symbols) and their randomized samples (open symbols).
}\end{figure}

\section{Assessment of long-range correlations in Martian and lunar topographies}\label{sec1}

In this section, we study the effect of correlations and their relevance on two observables, i.e., the total length of iso-height lines and the strength of the largest continent on Mars and the Moon. To this aim, we generate an uncorrelated version of martian and Lunar topographies by randomly shuffling the location of heights.  With "random reshuffling" we mean that all permutations of grid points are equally probable. We measure the total length of iso-height lines for both original and shuffled height profiles. As shown in Figure \ref{Fig9}, compared with its original data, this observable shows a very different behavior for the uncorrelated case on Mars. However, the spatially randomized topography on the Moon resembles the same behavior as its original structure with a very high overlap--- see Figure \ref{Fig9}.\\
The other quantity of interest is the strength of the largest continent as a function of altitude. Figure \ref{Fig10} shows our results for Mars and the Moon compared with their uncorrelated counterparts. We find that the largest gap in the strength of the largest continent for the Moon occurs at a height very close to its uncorrelated version, while for the martian topography they occur far apart from each other by $\sim 3$ km.  

These observations strongly suggest the existence of large-scale and non-random activities (such as tectonic and environmental activities which are missing on the moon) that have shaped the global topography of Mars in distant past.\bigskip

\begin{figure}
	\centerline{
		\includegraphics[width=0.8\textwidth]{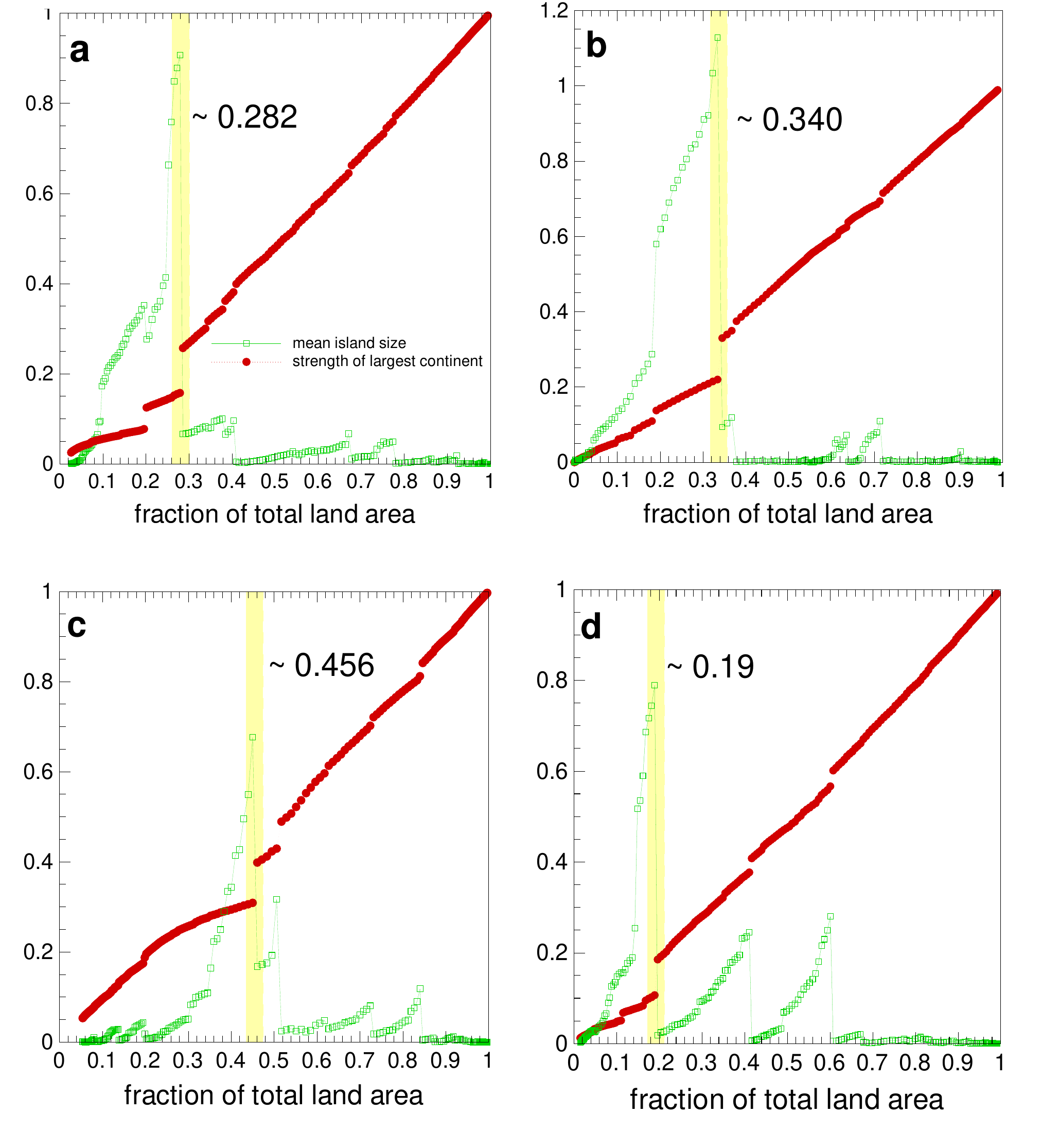}
	}
	\caption{Strength of the largest cluster (red solid circles) and mean cluster size (green open squares) as a function of the
		fraction of total land area for four different realizations of fBm surfaces with H = 0.5. The yellow shadow bars indicate the
		location of the threshold level associated with the largest jump in the order parameter and the divergence of mean cluster size.}
	\label{Fig11}
\end{figure}

\begin{figure}
\centerline{\includegraphics[width=0.8\textwidth]{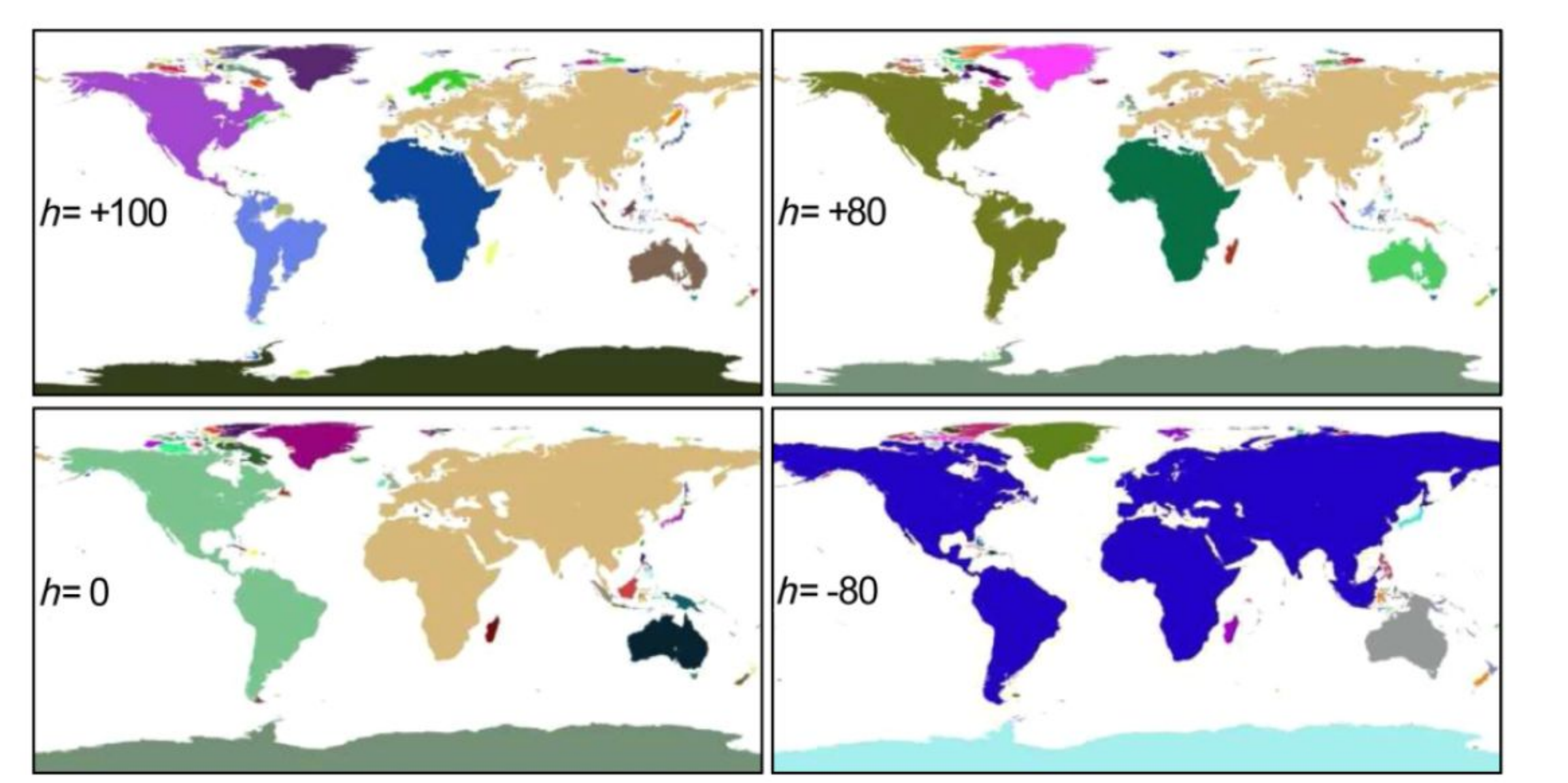}}
\caption{Schematic illustration of the continental aggregation by decreasing the sea level from top ($h=+100$ m) to bottom ($h=-80$ m). This shows a remarkable percolation transition at the present mean sea level around which the major parts of the landmass join together.}
\label{Fig12}
\end{figure}

%\bibliography{refs}

\end{document}